\documentclass[11pt, a4paper, onecolumn]{article}

\usepackage{fullpage}
\usepackage{graphicx}
\usepackage{wrapfig}
\usepackage[english]{babel}
\usepackage[center]{caption}
\usepackage{mathtools}
\usepackage{mathptmx}    
\usepackage{latexsym}           
\usepackage{url}
\usepackage{multirow}
\usepackage{amssymb}
\usepackage{multirow}
\usepackage{capt-of, blindtext}
\usepackage{pdfpages}
\usepackage{setspace}

\newcommand{\Keywords}[1]{\par\noindent {\bf {\em Keywords}: }#1 }

\begin{document}

\title{Are Quantum Models for Order Effects Quantum?\footnote{This work was supported by national funds through Funda\c{c}\~{a}o para a Ci\^{e}ncia e a Tecnologia (FCT) with reference UID/CEC/50021/2013 and through the PhD grant SFRH/BD/92391/2013. The funders had no role in study design, data collection and analysis, decision to publish, or preparation of the manuscript.}}

\author{Catarina Moreira\\ \small \texttt{catarina.p.moreira@tecnico.ulisboa.pt}\\
\and
Andreas Wichert\\ \small \texttt{andreas.wichert@tecnico.ulisboa.pt}
\and
\\Instituto Superior T\'{e}cnico / INESC-ID\\ Av. Professor Cavaco Silva, 2744-016 Porto Salvo, Portugal\\  
\\ \small The original publication is available at: International Journal of Theoretical Physics, Springer\\
 \small \text{\url{https://link.springer.com/article/10.1007/s10773-017-3424-5}}
}
\date{}

\maketitle							

\doublespacing	

\begin{abstract}

The application of principles of Quantum Mechanics in areas outside of physics has been getting increasing attention in the scientific community in an emergent discipline called Quantum Cognition. These principles have been applied to explain paradoxical situations that cannot be easily explained through classical theory. In quantum probability, events are characterised by a superposition state, which is represented by a state vector in a $N$-dimensional vector space. The probability of an event is given by the squared magnitude of the projection of this superposition state into the desired subspace. This geometric approach is very useful to explain paradoxical findings that involve order effects, but do we really need quantum principles for models that only involve projections? 

This work has two main goals. First, it is still not clear in the literature if a quantum projection model has any advantage towards a classical projection. We compared both models and concluded that the Quantum Projection model achieves the same results as its classical counterpart, because the quantum interference effects play no role in the computation of the probabilities. Second, it intends to propose an alternative relativistic interpretation for rotation parameters that are involved in both classical and quantum models. In the end, instead of interpreting these parameters as a similarity measure between questions, we propose that they emerge due to the lack of knowledge concerned with a personal basis state and also due to uncertainties towards the state of the world and towards the context of the questions.\\* 

\end{abstract}

\section{Introduction}

The application of principles of Quantum Mechanics in areas outside of physics has been getting increasing attention in the scientific community~\cite{Busemeyer12book}. These principles have been applied to explain paradoxical situations that cannot be easily explained through classical theory.  Quantum principles have been adopted in many different domains ranging from Cognitive Psychology~\cite{busemeyer06,Busemeyer09,Pothos13}, Economics~\cite{Khrennikov09eco,Haven13}, Biology~\cite{Asano12model,Basieva11,Asano15}, Information Retrieval~\cite{Bruza09quantum,Bruza13}, etc. 

One of these paradoxical situations is concerned with {\it order effects}. By order effects, we mean that the probability of, for instance, asking some question $A$ followed by question $B$ usually gives different results if we pose these questions in reverse order. In purely classical models, this poses a problem and cannot be directly explained. Since classical probability is based on set theory, this means that it is commutative, that is, for some hypothesis $H$ and two events $A$ and $B$: $Pr\left(~A \cap B~|~H~\right) = Pr\left(~B \cap A~|~H~\right)$. This commutativity poses great challenges to accommodate situations such as order effects, because in order to have $Pr\left(~H~|~A \cap B~\right) = Pr\left(~H~|~B \cap A~\right)$, then using Bayes Rule, one would need to satisfy the following relationship~\cite{Busemeyer11order}:
\begin{equation}
Pr\left(~H~|~A~\right) \cdot \frac{ Pr\left(~B~|~H \cap A~\right) }{ Pr\left(~B~|~A~\right)} = Pr\left(~H~|~A \cap B~\right) = Pr\left(~H~|~B \cap A~\right) = \frac{ Pr\left(~A~|~H \cap B~\right)}{ Pr\left(~A~|~B~\right)} \cdot Pr\left(~H~|~B~\right).
\end{equation}
To accommodate these paradoxical findings, the literature turned to quantum probability to explain these scenarios. Quantum probability models provide many advantages towards their classical counterparts~\cite{Busemeyer15comparison}. They can represent events in vector spaces through a superposition state, which comprises the occurrence of all events at the same time. In quantum mechanics, the superposition principle refers to the property that particles must be in an indefinite state. That is, a particle can be in different states at the same time. Under a psychological point of view, a quantum superposition can be related to the feeling of confusion, uncertainty or ambiguity~\cite{Busemeyer12book}. The vector space representation does not obey the distributive axiom of Boolean logic and to the law of total probability. This enables the construction of more general models that can mathematically explain cognitive phenomena such as order effects~\cite{Busemeyer09,Khrennikov09sure}.

One of the quantum approaches that is highly used to explain order effects is the quantum projection model (also known as a quantum geometric model)~\cite{Pothos13,Trueblood14}. In this approach, we start to represent a person's beliefs in a $N$-dimensional superposition state (for the case of binary questions, N = 2). To model a sequence of answers, we start by projecting this superposition state into the basis with the desired answer. From this basis, we perform a second projection to another basis, which represents the desired answer for the second question, and corresponds to a rotation by $\phi$ radians of the initial basis state. The final probability is given by performing the squared magnitude of the multiplication of all these projections. Since projections are given by matrices and matrix multiplication is non-commutative, then order effects can be naturally explained under this framework. 

The questions that we address in this work are the following. Given that the quantum approach consists in a geometric model that performs projections, can we obtain the same results using a classical approach also based on projections? Is quantum theory really necessary and advantageous to explain paradoxical findings such as order effects? And, what makes this quantum projection approach {\it quantum}, since no complex probability amplitudes are being used in the model~\cite{Busemeyer12book,Wang14}? From where do the quantum interferences arise from and what do they mean under this context? 

In this work, we pretend to make a discussion about these questions. We also propose an alternative interpretation for the parameters that are involved in these geometric projection models that can be applied to both classical and quantum models. Current literature interprets these parameters as similarities between questions. Under the proposed Relativistic Interpretation, these parameters emerge due to the lack of knowledge concerned with a personal basis state and also due to uncertainties towards the state of the world and towards the context of the questions. So, with a relativistic interpretation of parameters, we can give to both classical and quantum approaches an interpretation for the rotation of the basis vectors and why this rotation is necessary under a cognitive point of view.

In the end, we argue that the application of the classical and quantum models should be based on Occam's Razor: in the presence of two competing hypothesis, the one that has the fewest assumptions (or the one that is simpler) should be chosen. This depends much on the problem and what knowledge we want to extract from it. If we are mainly focused on a mathematical approach that can perform predictions for order effects, abstracting the model from any interpretations or theories, which are intrinsic to the problem, then the classical approach is the one that makes the fewer assumptions and should be applied. If, on the other hand, we want to make a model that leverages on a more general theory to explain its predictions, then the quantum model is more indicated. 

This work is organised as follows. In Section~\ref{sec:poll}, we present the results of a study from Moore (2002)~\cite{Moore02} in which the author collected public opinions by making two consecutive questions regarding important people and relevant issues. The results collected in this gallup poll showed the occurrence of several types of order effects. In Section~\ref{sec:qpa}, we explain how the quantum projection model works and how it can explain the results reported by~\cite{Moore02}. In Section~\ref{sec:riqp}, we present an alternative interpretation for the parameters involved in the quantum projection approach that is general enough to be applied to both quantum and classical models. In Section~\ref{sec:cpa}, we show how a classical model can obtain the same results as a quantum approach and also explain the paradoxical findings of order effects. In the end, we perform a discussion in which situations one should prefer the classical or the quantum model. Finally, we end this work with Section~\ref{sec:conc}, in which we present some final discussions and the main conclusions of this work.

\section{The Gallup Poll Problem}\label{sec:poll}

One example of question order effects that has been widely reported over the literature corresponds to the work of~\cite{Moore02}, where the author collected public opinions between two important people: Bill Clinton and Al Gore. In this poll, half of the participants were asked if they thought that Bill Clinton was a honest and trustworthy person and next they were asked the same about Al Gore. The other half of the participants were asked the same questions, but in {\it reversed order}. In the end, there was a total of $1002$ respondents. 

Results showed that there was a big gap in a non-comparative context (the first questions) and a small gap in a comparative context (the second question). Asking about Clinton first, made the probability of the second question increase. On the other hand, asking about Al Gore first, made the probability of the second question decrease. This phenomena is usually referred to \emph{the assimilation effect} and pure classical probability models cannot explain it, because they are based on set theory and, consequently, they are commutative. Table~\ref{tab:moores_data1} summarises the results obtained in the work of~\cite{Moore02}.

\begin{table}[h!]
\centering
\begin{tabular}{l | c | c | c }	
~							& \textbf{Clinton-First}		& \textbf{Gore-First}		& \textbf{Differences}	\\
\hline
Pr After First Question		& $Pr(C) = 50\%$				& $Pr(G) = 68\%$				& $ = 18\%$ (non-comparative) \\
Pr After Second Question	& $Pr(G | C) = 57\%$			& $Pr(C | G) = 60\%$			& $ = 3\%$  (comparative)	\\
Effect					& + 7\%						&	- 8\%					& Assimilation Effect	\\
\hline
\end{tabular}
\caption{Summary of the results obtained in the work of~\cite{Moore02} for the Clinton-Gore Poll, showing an Assimilation Effect}
\label{tab:moores_data1}
\end{table}

In the same work,~\cite{Moore02} also reported the same experiment, but using different politicians in the questions. The same questions as above were asked, but regarding the honesty and trustworthiness of Gingrich and Dole. The questions were posed in different orders. The total amount of respondents in this experiment was $1015$. Table~\ref{tab:moores_data2} summarises the results.

Opposite to the assimilation effect, asking about Gingrich first, made the probability of the question regarding Dole decrease. On the other hand, asking about Dole first made the probability of the question regarding Gingrich increase. This phenomena is usually referred to \emph{the Contrast effect}. Table~\ref{tab:moores_data2} summarises the results obtained in the work of~\cite{Moore02}.

\begin{table}[h!]
\centering
\begin{tabular}{l | c | c | c }	
~							& \textbf{Gingrich-First}		& \textbf{Dole-First}		& \textbf{Differences}	\\
\hline
Pr After First Question		& $Pr(G) = 41 \%$				& $Pr(D) = 60 \%$				& $ = 19\%$ (non-comparative) \\
Pr After Second Question	& $Pr(D | G) = 33 \%$			& $Pr(G | D) = 64 \%$			& $ = 31\%$  (comparative)	\\
Effect					& - 8\%						& + 4\%						& Contrast Effect	\\
\hline
\end{tabular}
\caption{Summary of the results obtained in the work of~\cite{Moore02} for the Gingrich-Dole Poll, showing a Contrast Effect.}
\label{tab:moores_data2}
\end{table}

The third poll reported in the work of~\cite{Moore02} corresponds to a set of questions concerned with racial hostility. In a group of 1004 respondents, two questions were asked. One was: {\it Do you think that only a few white people dislike blacks, many white people dislike blacks, or almost all white people dislike blacks?}. The other question was the same, but about black people. These questions were posed in sequence and in different orders. Table~\ref{tab:moores_data3} summarises the results. 

In this case, the order how the questions were posed did not matter, since they both contributed to an increase of the probability of the second question. This phenomena is usually referred to \emph{the additive effect}.

\begin{table}[h!]
\centering
\begin{tabular}{l | c | c | c }	
~						& \textbf{White People First}		& \textbf{Black People First}	& \textbf{Differences}	\\
\hline
Pr After First Question		& $Pr(G) = 41 \%$				& $Pr(D) = 46 \%$				& $ = 5\%$ (non-comparative) \\
Pr After Second Question	& $Pr(D | G) = 53 \%$			& $Pr(G | D) = 56 \%$			& $ = 3\%$  (comparative)	\\
Effect					& + 12\%						& + 10\%						& Additive Effect	\\
\hline
\end{tabular}
\caption{Summary of the results obtained in the work of~\cite{Moore02}. The table reports the probability of answering $All$ or $Many$ to the questions. The results show the occurrence of an Additive Effect.}
\label{tab:moores_data3}
\end{table}

The last example in the work of~\cite{Moore02} is concerned with a poll about Peter Rose and Joe Jackson. Again, a set of $1061$ respondents was gathered and two questions were posed in sequence. The questions were {\it do you think Peter Rose / Shoeless Joe Jackson should or should not be eligible for admission to the Hall of Fame?}. These questions were performed in different orders and the results obtained are summarized in Table~\ref{tab:moores_data4}.

In this last example, the order how the questions were posed did not matter, since they both contributed to a decrease of the probability of the second question. This phenomena is usually referred to \emph{the subtractive effect}.

\begin{table}[h!]
\centering
\begin{tabular}{l | c | c | c }	
~						& \textbf{Peter Rose First}		& \textbf{Shoeless Joe Jackson First}	& \textbf{Differences}	\\
\hline
Pr After First Question		& $Pr(G) = 64 \%$				& $Pr(D) = 45 \%$				& $ = 19 \%$ (non-comparative) \\
Pr After Second Question	& $Pr(D | G) = 52 \%$			& $Pr(G | D) = 33 \%$			& $ = 19 \%$  (comparative)	\\
Effect					& - 12\%						& - 12\%						& Subtractive Effect	\\
\hline
\end{tabular}
\caption{Summary of the results obtained in the work of~\cite{Moore02} for the Rose-Jacjson Poll, showing a Subtractive Effect.}
\label{tab:moores_data4}
\end{table}

All data reported in the work of Moore (2002)~\cite{Moore02} constitute a violation of order effects. By order effects, we mean the probability of, for instance, asking some question $A$ followed by question $B$ usually gives different results if we pose the questions in reverse order. One way to approach this problem is through Quantum Cognition Models. Quantum cognition is a research field that aims to explain paradoxical findings (such as order effects) using the laws of quantum probability theory. The models provide several advantages towards its classical counterparts. They can model events in a superposition state, which is a vector that comprises the occurrence of all possible events. They enable events in superposition to interfere with each other, this way disturbing the final probability outcome. These quantum interference effects do not exist in a classical setting and constitute the major advantage of these models, since we can use these interference effects to accommodate the paradoxical findings.


\section{A Quantum Approach for Order Effectsl}\label{sec:qpa}

In the Quantum Projection Model~\cite{Busemeyer12book,Pothos13,Trueblood14}, a state is represented by a unit vector in a $k$-dimensional complex vector space. A quantum superposition state is represented by a vector $| \psi \rangle = \alpha_0 | 0 \rangle + \alpha_1 | 1 \rangle + \dots + \alpha_{k-1} | k-1 \rangle$, where $\alpha_0, \dots, \alpha_{k-1}$ are quantum amplitudes and the sum of their squared magnitude must sum to $1$, $\sum_{i=0}^{k-1}  \left| \alpha_i \right|^2 = 1$. In the case of the Gallup Poll presented in Section~\ref{sec:poll}, the quantum states are binary, since they correspond to a person's $yes / no$ answer, that is, we would represent a superposition state $| S \rangle$ regarding the answer to Clinton's honesty and trustworthiness person as
\begin{equation}
| S \rangle = s_0 |Cy \rangle + s_1 | Cn \rangle,
\end{equation}
where $Cn$ and $Cy$ correspond to the answers $no$ or $yes$, respectively to Clinton's honesty question. The variables $s_0$ and $s_1$ represent complex quantum probability amplitudes, which represent the initial beliefs of a participant, before answering the question.

\begin{figure}[h!]
\centering
\includegraphics[scale=0.5]{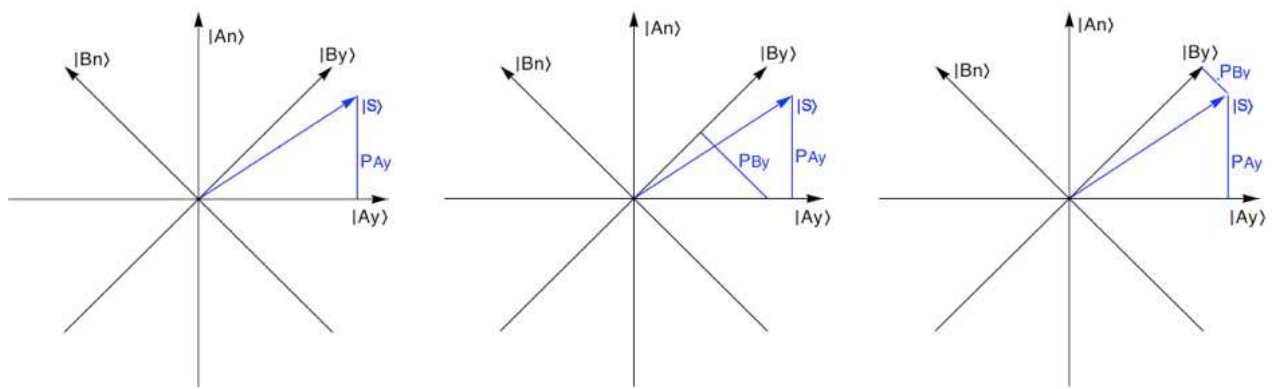}
\caption{Example of the application of the quantum projection approach for a sequece of two binary questions $A$ and $B$. We start in a superposition state and project this state into the $yes$ basis of question $A$ (left). Then, starting in this basis, we project into the basis corresponding to the answer $yes$ of question $B$ (center). We can then have a different result if we reverse the order the projections (right).}
\label{fig:proj}
\end{figure}

Consider Figure~\ref{fig:proj} where it is represented two sets of basis states: $\{ Ay, An \}$ and $\{ By, Bn \}$. Generally speaking, one can look at these basis as the representation of two $yes/no$ questions. Basis state $Ay$ corresponds to a $yes$ answer to question $A$ and is given by the state vector $| Ay \rangle = [~\begin{matrix} 1 & 0 \end{matrix}~]$ and $| An \rangle = [~\begin{matrix} 0 & 1 \end{matrix}~]$ corresponds to a $no$ answer to question $A$. In the same way, for the $yes/no$ answers of question $B$, we can represent the states as a {\it rotation} of questions' $A$ basis states. When an answer is given, we project the superposition state $| S \rangle$ into the basis state corresponding to the desired answer. In Figure~\ref{fig:proj} (right), we perform two sets of projections: (1) starting in the superposition state $|S \rangle$, we first make a projection that is orthogonal to the basis state representing the $yes$ answer of question $B$ giving rise to the projector $P_{By}$ and (2) from the basis state $By$ we perform another orthogonal projection to the $yes$ basis vector of question $A$, resulting in projector $P_{Ay}$. In Figure~\ref{fig:proj} (center), we perform the same projections, but in reverse order:  (1) starting in the superposition state $|S \rangle$, we first make a projection that is orthogonal to the basis state representing the $yes$ answer of question $A$, $P_{Ay}$, and then (2) from the basis state $Ay$ we perform another orthogonal projection to the $yes$ basis vector of question $B$, $P_{By}$. 

In the end, the final probability of answering $yes$ to $B$ given that it was previously answered $yes$ to question $A$ is given by the squared magnitude of the sequence of these projections. We can obtain a different probability value by making the inverse sequence of questions. The probabilities computed using the Quantum Projection approach give different results, matching the experimental findings of Moore (2002)~\cite{Moore02}, in which it is shown that the order of how the questions are posed matter and have an impact over the results.  
\begin{equation}
Pr( B = yes ) = \left| \left| PAy~PBy~| S \rangle \right| \right|^2 + \left| \left| PAn~PBy~| S \rangle \right| \right|^2 \neq \left| \left| PBy~PAy~| S \rangle \right| \right|^2 + \left| \left| PBy~PAn~| S \rangle \right| \right|^2 
\end{equation}

\subsection{The Quantum Projection Model}

For a 2-dimensional decision scenario the projection model can be described in the following points:
\begin{itemize}
\item Start by defining two orthonormal basis vectors for the questions, which implies that $\langle An | Ay \rangle=0$. One set of basis vectors that is commonly chosen is
\[ |Ay \rangle = \left[ \begin{matrix} 1 \\ 0 \end{matrix} \right] ~~~~~~~~ |An \rangle = \left[ \begin{matrix} 0 \\ 1 \end{matrix} \right], \] 
the other set of vectors corresponds to a rotation of the above basis vectors by $\phi$ radians. The new basis can be found by multiplying a rotation matrix $R_{\phi}$ with each of the basis vectors
\begin{equation}
|By \rangle = R_{\phi} Ay = 
\left[ \begin{matrix} 
	Cos\left( \phi \right) \\ 
	Sin\left( \phi \right) 
	\end{matrix} 
\right]~~~~~~~~ 
|Bn \rangle = R_{\phi} An = 
\left[ \begin{matrix} 
	-Sin\left( \phi \right) \\ 
	Cos\left( \phi \right) 
	\end{matrix} 
\right]~~~~~~~~
R_\phi =  
\left[ \begin{matrix} 
	Cos\left( \phi \right)	& -Sin\left( \phi \right) \\ 
	Sin\left( \phi \right)	& Cos\left( \phi \right) 
\end{matrix} \right].
\end{equation} 
\item Then, we define a superposition vector $| S \rangle$, which comprises a person's beliefs (or features) about some object. In this case, it corresponds to a superposition of possible answers to a given question: to answering {\it No} (An) or {\it Yes} (Ay). Although the model can be generalized for $N$ random variables, we will describe the model for two random variables, which is what we need to describe the different order effects found in the work of~\cite{Moore02}.
\begin{equation}
| S \rangle = \sqrt{s_0}~e^{i\theta_0} | Ay \rangle + \sqrt{s_1}~e^{i\theta_1} | An \rangle \text{~~~~~~~~~~such that,~~}  \sum_{j=0}^{N=1} \left| \sqrt{s_j}~e^{i\theta_j} \right|^2 = 1. 
\end{equation}
The variables $s_0$ and $s_1$ represent quantum probability amplitudes and the variables $\theta_0$ and $\theta_1$ correspond to their respective phase.

\item When we ask question $A$ first, this corresponds to a projection of the belief state $|S \rangle$ onto the subspace with the desired answer. In Figure~\ref{fig:proj} (left), this vector is being projected into the subspace $Ay$, which corresponds to the answer $yes$ of question $A$. This projection produces the state $P_{Ay} | S \rangle$, which is given by
\begin{equation}
P_{Ay} = | Ay \rangle \langle Ay | = \left[ \begin{matrix} 1	& 0 \\
									  0	& 0 \end{matrix} \right] ~~~~~~~~ P_{Ay} | S \rangle = \left[ \begin{matrix} \sqrt{s_0}~e^{i\theta_0} \\ 0 \end{matrix} \right] 
\end{equation}
\item The probability of answering $yes$ to the first question is given by the squared magnitude of the projected state:
\begin{equation}
\left| \left| P_{Ay} | S \rangle \right| \right|^2 = \left| \left| \left[ \begin{matrix} s_0~e^{i\theta_0} \\ 0 \end{matrix} \right] \right| \right|^2 = \sqrt{s_0}~e^{i\theta_0} \cdot \left(\sqrt{s_0}~e^{i\theta_0} \right)^*= s_0
\end{equation} 

\item When we ask question $B$ in the first place, this corresponds to a projection operator $P_{By}$ that will project the superposition state $S$ into the $yes$ basis of question $B$. For this, we need to define the projector $P_{By}$ and the rotation of the superposition state $S_R$ as
\begin{equation}
P_{By} = | By \rangle \langle By | = 
	\left[ \begin{matrix} 
		Cos^2\left( \phi \right) & Cos\left( \phi \right) Sin\left( \phi \right) \\  
		Sin\left( \phi \right) Cos\left( \phi \right) & Sin^2\left( \phi \right)  
		\end{matrix} \right] 
\end{equation}
So, $P_{By} | S \rangle$ is given by,
\begin{equation}
P_{By} = \left( \left| Cos\left( \phi \right) \right|^2 + \left| Sin\left( \phi \right) \right|^2  \right) \left| \sqrt{s_0}~e^{i\theta_0}~Cos\left( \phi \right) + \sqrt{s_1}~e^{i\theta_1}~Sin\left( \phi \right)  \right|^2
\end{equation}
			
\item In order to compute the probability of the sequence of answers $Ay \rightarrow By$, we need to compute the squared magnitude of their sequence of projections. That is, we need to compute $\left| \left| P_{By} P_{A_y} | S \rangle \right| \right|^2$. This sequence of projections is illustrated in Figure~\ref{fig:proj} (left) and corresponds answering $yes$ to question $A$ and $yes$ to question $B$, respectively.
\begin{equation}
Pr( By Ay ) = \left| \left| P_{By} P_{Ay} | S \rangle \right| \right|^2 = Cos^2\left( \phi \right)  
	\left|  \sqrt{s_0}~e^{i \theta_0} Cos\left( \phi \right) +
		 \sqrt{s_1}~e^{i \theta_1} Sin\left( \phi \right)   \right|^2
\end{equation}

\item In the same way, we can compute the sequence of answers $Bn$ and $Ay$, representing the answer $yes$ to question $A$ and $no$ to question $B$.
\begin{equation}
Pr( Bn Ay ) =  \left| \left| P_{Bn} P_{Ay} | S \rangle \right| \right|^2  =  Sin^2\left(  \phi \right) 
	\left| \sqrt{s_1}~e^{i \theta_1} Cos\left( \phi \right) - 
		\sqrt{s_0}~e^{i \theta_0} Sin \left( \phi  \right)  \right|^2
\end{equation}

\item The final probability of $A$ being $yes$, $Pr(Ay)$, is given by the sum: $Pr( B \rightarrow Ay) = Pr( By Ay ) + Pr( Bn Ay )$.
\begin{equation}
\begin{split}
Pr( B \rightarrow Ay) =Pr( By Ay ) + Pr( Bn Ay )=  
Sin^2\left( \phi \right) \left| Cos\left( \phi \right) \sqrt{s_1}~e^{ i \theta_1} - 
    \sqrt{s_0}~e^{ i \theta_0} Sin\left( \phi \right) \right|^2 + \\
 Cos^2\left( \phi \right) \left| Cos\left( \phi \right) \sqrt{s_0}~e^{ i \theta_0} + \sqrt{s_1}~e^{ i \theta_1} Sin\left( \phi \right) \right|^2~~~
\end{split}
\label{eq:ay}
 \end{equation}

\item If, however, we wanted to compute the probability of $By$ in the second question, that is, the probability of answering $yes$ to $B$ after answering question $A$, then we would have to compute the sequence of projections:

\begin{equation}
 Pr( Ay By ) = \left| \left| P_{Ay}~P_{By} | S \rangle \right| \right|^2  =  \left| \sqrt{s_0}~e^{i \theta 0} \right|^2~Cos^2\left( \phi \right) = \left|~s_0~\right| Cos^2\left( \phi \right)
\end{equation}

\item The probability of answering $no$ to the first question $A$ and $yes$ to the second question $B$ is given by
\begin{equation}
 Pr( An By ) = \left| \left| P_{An}~P_{By} | S \rangle \right| \right|^2  =  \left| \sqrt{s_1}~e^{i \theta 1} \right|^2~Sin^2\left( \phi \right) = \left| s_1 \right|Sin\left( \phi \right)
\end{equation}

\item The final probability of answering $yes$ to question $B$ followed by question $A$ is given by the sum of the probabilities $Pr( A \rightarrow By ) = Pr( AnBy) + Pr( Ay By )$
\begin{equation}
Pr( A \rightarrow By )  =  \left| \sqrt{s_0}~e^{i \theta 0} \right|^2~Cos^2\left( \phi \right)+ \left| \sqrt{s_1}~e^{i \theta 1} \right|^2  Sin^2\left( \phi \right)  = \left| s_0 \right|~Cos^2\left( \phi \right)+  \left| s_1 \right|~Sin^2\left( \phi \right)  
\label{eq:second_quest_final}
\end{equation}

\item In the end, setting the parameters as suggested in~\cite{Busemeyer12book}, that is $\theta = \pi/4$, $s_0 = 0.7$ and $s_1 = 0.3$, leads to a big gap in the non-comparative context and a small gap for the comparative context, simulating the {\it Assimilation Effect} presented in Table~\ref{tab:moores_data1}.
\[ Pr( Ay ) = 0.7~~~~~Pr( By ) = 0.96~~~~~Pr( A \rightarrow~By ) = Pr( B \rightarrow~Ay )  = 0.5 \]
\end{itemize}

\subsection{Discussion of the Quantum Projection Model}

\begin{figure}[b!]
	\parbox{.45\linewidth}{
	\centering
	\includegraphics[scale=0.55]{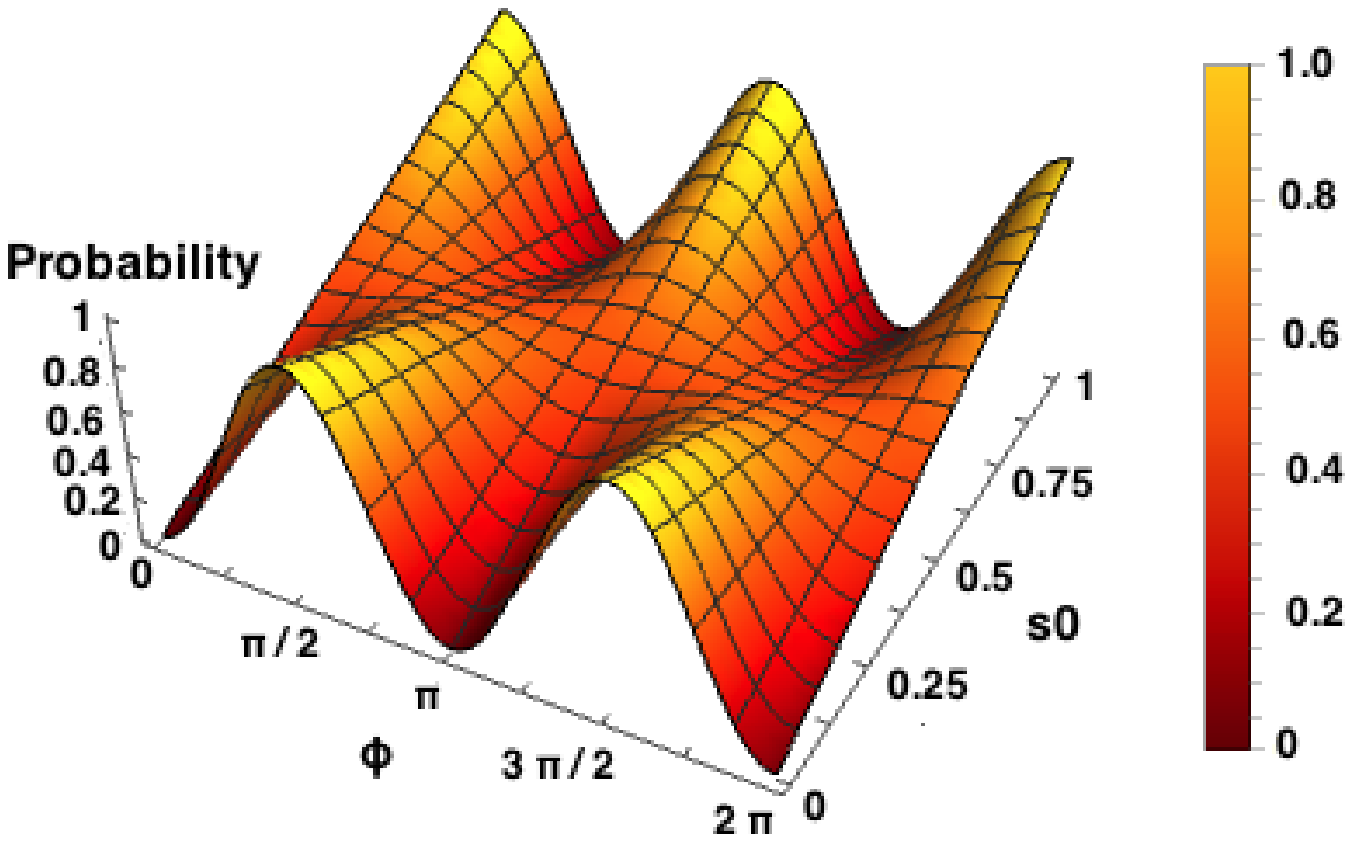}
	\caption{Relation between the rotation parameter $\phi$ and the quantum probability amplitude $s_0$ of Equation~\ref{eq:second_quest_final}. The amplitude $s_1$ was set to $s_1= 1 - s_0$. We can simulate several order effects by varying the parameter $\phi$. }
	\label{fig:var_phi_s0B}
	}
	\hfill
	\parbox{.45\linewidth}{
	\centering
	\includegraphics[scale=0.55]{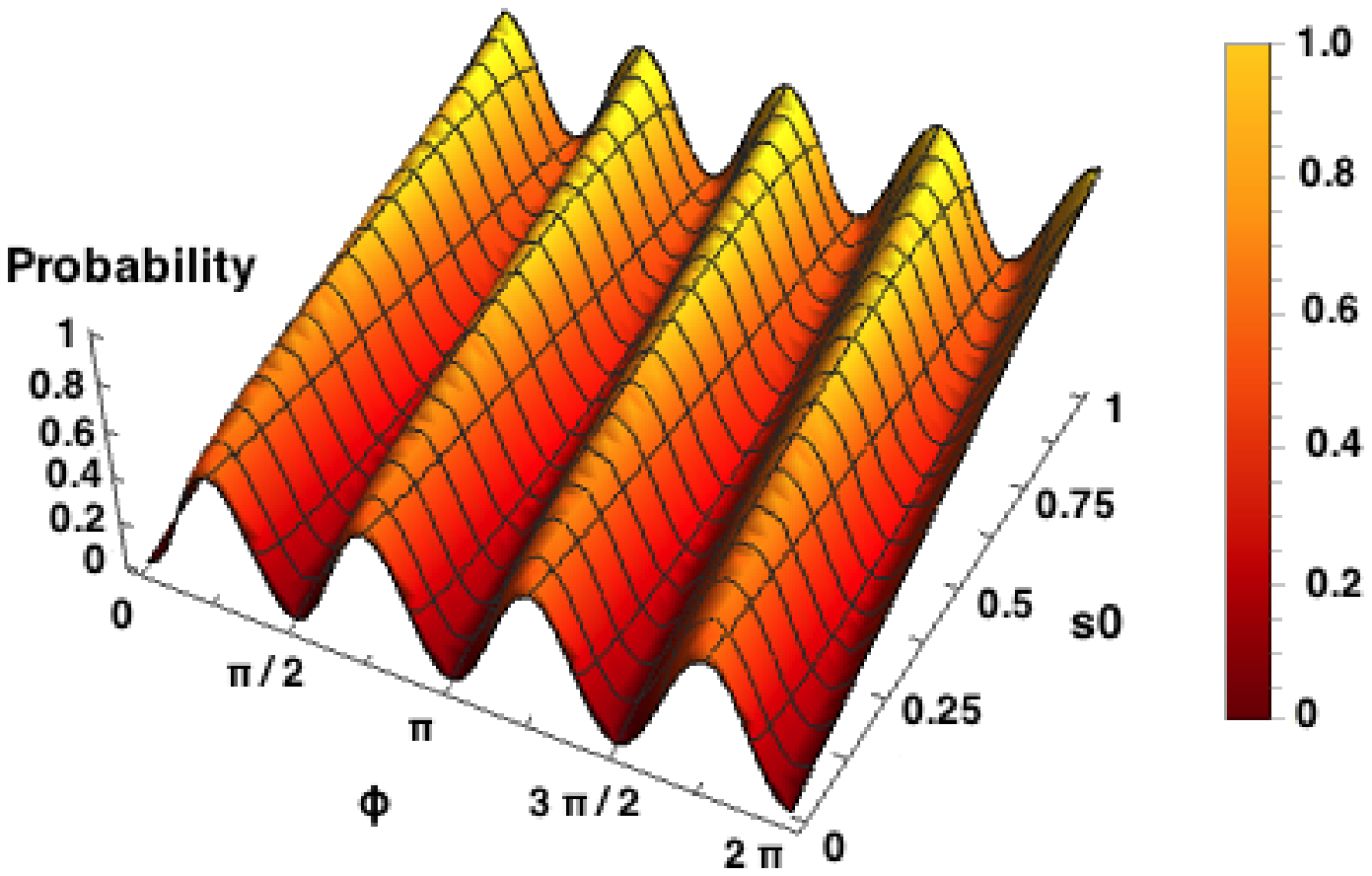}
	\caption{Relation between the rotation parameter $\phi$ and the quantum probability amplitude $s_0$ of Equation~\ref{eq:ay}. The amplitude $s_1$ was set to $s_1= 1 - s_0$. We can simulate several order effects by varying the parameter $\phi$. }
	\label{fig:s01}
	}
\end{figure}

In the quantum projection model, it is interesting to notice that in Equation~\ref{eq:second_quest_final}, the probability of answering $yes$ to question $B$ followed by question $A$, does not depend on the quantum parameter $\theta$. This means that the quantum model collapses to its classical counterpart, since it depends only on the rotation parameter $\phi$ and on the initial beliefs (given by $s_0$ and $s_1$). A deeper analysis of Equation~\ref{eq:second_quest_final} shows more information. If we fix the rotation parameter $\phi$ to $\pi/4$ and $s_1 = 1 - s_0$ (as suggested in the example in~\cite{Busemeyer12book}), then we can conclude that varying the initial belief state $s_0$ plays no role in the outcome of the final probabilities. When making a sequence of questions, only the rotation parameter $\phi$ can change the outcome. In another analysis, we tested how the function would evolve if we varied the rotation parameter $\phi$ (between $0$ and $2 \pi$) and the initial belief state $s_0$ (between 0 and 1). The outcome is Figures~\ref{fig:var_phi_s0B} and~\ref{fig:s01}, in which one can conclude that it is possible to simulate the several order effects reported in the work of Moore (2002)~\cite{Moore02} by setting $s_0$ and varying the rotation parameter $\phi$.

In the same way, we performed a similar analysis to Equation~\ref{eq:ay}, which corresponds answering $yes$ to the sequence of questions $B \rightarrow Ay$. The difference is that Equation~\ref{eq:ay} does depend on the quantum interference parameter $\theta$. By fixing the rotation parameter $\phi$ and the initial belief state $s_0$ with the quantum interference parameter $\theta$,  this analysis showed similar results to the previous question: when we fix the rotation parameter, the function becomes constant and both initial state $s_0$ and quantum interference term $\theta$ play no role in the final probability outcome. This means that when computing these probabilities, only the rotation parameter will have a direct impact in the calculations. Moreover, if we fix the quantum interference term $\theta = 0$ (as suggested in the example contained in~\cite{Busemeyer12book}), then we can see that varying the rotation parameter $\phi$ also enables the possibility of representing the different types of order effects reported in~\cite{Moore02}: {\it assimilation}, {\it contrast}, {\it additive} and {\it subtractive} effects. In the next section, we will address these rotation parameters more closely and propose and alternative interpretation under the scope of quantum cognition.

\section{The Relativist Interpretation of Parameters}\label{sec:riqp}

In the book of~\cite{Busemeyer12book}, quantum parameters in the Quantum Projection Model represent the similarity between questions. This parameter represents the angle of the inner product between the two projections. In other works of the literature, the quantum interference parameter represents not only the similarity between two random variables (through their inner product), but also the semantical relationship between them~\cite{Moreira16,Moreira16faces}.

However, under the Quantum Projection Model, we have seen in the previous section that the quantum interference parameters play no role in the computation of the final probabilities. So, where do the quantum interference effects come from in order to accommodate the violations of order effects? One could argue that this interference comes from the rotation of the basis vectors by $\phi$ radians. In the previous section, the rotation operator $R$ changed the basis state $A_x$ into $B_x$ by applying a rotation of $\phi$ radians. Under a cognitive point of view, this rotation is used to model the quantum interference effects and corresponds to the change of a person's mental beliefs~\cite{busemeyer12question,Wang14}. However, this interpretation is independent from a quantum perspective and it holds for pure classical projection models as we will show in Section~\ref{sec:cpa}.

In this section, we present an alternative interpretation of quantum interference parameters for this projection-based approach for order effects that can accommodate both classical and quantum projection approaches for order effects. We call it {\it The Relativistic Interpretation of Parameters}.

Under the Relativistic Interpretation of Parameters, when we pose a question to different people, each person will represent its preferred answer in a $N$-dimensional vector space. For the case of binary questions, this representation is performed in a $2$-dimensional psychological space. However, the individual is not aware in which basis he is making this representation. Each individual person has its own basis, but it is possible to represent different beliefs from different individuals by performing a rotation of the basis state by $\phi$ radians.

\begin{figure}[h!]
\centering
\includegraphics[scale=0.45]{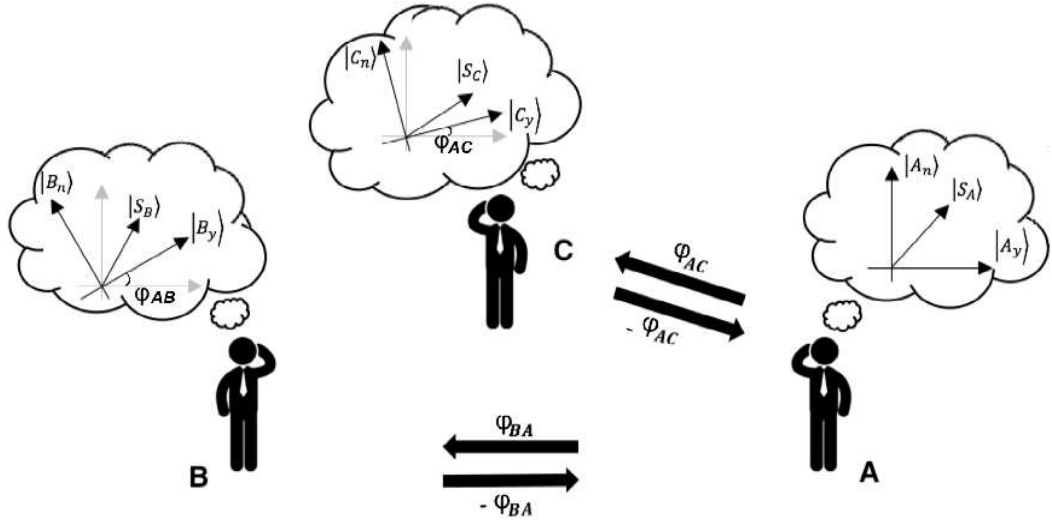}
\caption{Example of the Relativistic Interpretation of Quantum Parameters. Each person reasons according to a N-dimensional personal basis state without being aware of it. The representation of the beliefs between different people consists in rotating the personal belief state by $\phi$ radians. }
\label{fig:rel}
\end{figure}

Take as example Figure~\ref{fig:rel}. There are three individuals $A$, $B$ and $C$ to whom it is posed the same binary question. Each person will represent their answer in their own 2-dimensional psychological space, without knowing in which basis they are representing their beliefs. However, since we are assuming that each person has its own vector space, then one can describe the beliefs of each individual in terms of another by performing a rotation of their basis by some $\phi$ radians. For instance, assuming that individual $A$ is in the common $| 0 \rangle$ / $| 1 \rangle$ basis, then, individual's $C$ beliefs are described, under person's $A$ perspective, as a rotation of $\phi_{A,C}$ radians. In the same way, person's $A$ beliefs can be described by person's $C$ point of view by performing the inverse rotation, that is, by  rotating the basis $- \phi_{A,C}$ radians or $\phi_{C,A}$ radians. The same line of thought is applied for person $B$. 

In the end, quantum interferences arise in quantum cognition due to this {\it lack of
knowledge} regarding each person's own basis states and due to {\it uncertainties towards the state world}.

In the next section, we will perform a discussion of whether or not we need a Quantum Approach to model order effects.

\section{Do We Need Quantum Theory for Order Effects?}\label{sec:cpa}

So far, we presented the problem of order effects in which the probability of a sequence of events is different if we switch the order of that sequence. This cannot be simulated by pure classical probabilistic models, because classical probability theory is based on set theory and, consequently, events commute.

We also presented a quantum model that is widely used in the literature and can account for order effects in a general and natural way. However, we showed that the quantum interferences do not play any role in this quantum projection model and only the rotation of the basis vectors is necessary to accommodate paradoxes derived from order effects. At this point, the reader might be thinking: {\it Do we really need quantum theory to explain order effects? Can we achieve the same results using a classical projection counterpart of the quantum model?} The answers to these questions are {\it no} and {\it yes}, respectively.

One can argue that the main difference between the quantum approach towards a classical projection model is that in the first it is used complex probability amplitudes and in the later it is used real numbers. One could argue that what makes the model quantum is the usage of these quantum amplitudes, which in turn generate quantum interference effects, which can accommodate several paradoxical findings reported in the literature~\cite{Tversky74,Tversky83Uncertainty,Tversky92,Birnbaum08}. However, that does not hold in the quantum projection model, since quantum interferences end up playing no role in the computation of the probabilities. Moreover, even if the quantum interference terms did matter, then the complexity of the quantum projection model would increase, since we would require an additional $2^N$ free quantum interference parameters for binary questions that would need to be fit. For the case of $M$ possible answers, the complexity will grow to $M^N$, where $N$ is the number of questions.

 

\subsection{A Classical Approach for Order Effects}

The classical projection approach works just like the previously described quantum model with the difference that real numbers are used instead of quantum probability amplitudes. The model works in a N-dimensional vector space, however, in order to simulate the results obtained in the Clinton / Gore experiment from the work of Moore (2002), we will describe the model for a 2-dimensional decision scenario:

\begin{itemize}
\item Start by defining two orthonormal basis vectors for the questions. One set of basis vectors that is commonly chosen is
\[ |Ay \rangle = \left[ \begin{matrix} 1 \\ 0 \end{matrix} \right] ~~~~~~~~ |An \rangle = \left[ \begin{matrix} 0 \\ 1 \end{matrix} \right], \] 
the other set of vectors corresponds to a rotation of the above basis vectors by $\phi$ radians. The new basis can be found by multiplying a rotation matrix $R_{\phi}$ with each of the basis vectors
\begin{equation}
|By \rangle = R_{\phi} Ay = 
\left[ \begin{matrix} 
	Cos\left( \phi \right) \\ 
	Sin\left( \phi \right) 
	\end{matrix} 
\right]~~~~~~~~ 
|Bn \rangle = R_{\phi} An = 
\left[ \begin{matrix} 
	-Sin\left( \phi \right) \\ 
	Cos\left( \phi \right) 
	\end{matrix} 
\right]~~~~~~~~
R_\phi =  
\left[ \begin{matrix} 
	Cos\left( \phi \right)	& -Sin\left( \phi \right) \\ 
	Sin\left( \phi \right)	& Cos\left( \phi \right) 
\end{matrix} \right].
\end{equation} 
\item Since we are in a Euclidean space, we can define a vector $| S \rangle$, where $s_0$ and $s_1$ are variables representing real numbers.
\begin{equation}
| S \rangle = \sqrt{s_0} | Ay \rangle + \sqrt{s_1} | An \rangle \text{~~~~~such that~~~~~} \sum_{i=0}^{N=1} \left| \sqrt{s_i} \right|^2 = 1.
\end{equation}

\item The probability of answering $yes$ to the first question $A$ corresponds to the squared magnitude of the projection of vector $S$ into the $yes$ basis of $A$:
\begin{equation}
P_{Ay} = | Ay \rangle \langle Ay | = \left[ \begin{matrix} 1	& 0 \\
									  0	& 0 \end{matrix} \right]  ~~~~~~~ Pr\left( Ay \right) = \left| \left| P_{Ay} | S \rangle \right| \right|^2 = \left| s_0 \right|
\end{equation}
\item On the other hand, if we want to compute the probability of posing question $B$ first, then the same paradigm applies. We start with vector $S$ and then we project this state into the $yes$ basis of question $B$:
\begin{equation}
P_{By} = | By \rangle \langle By | = \left[ \begin{matrix} Cos^2\left( \phi \right) & Cos\left( \phi \right) Sin\left( \phi \right) \\  Sin\left( \phi \right) Cos\left( \phi \right) & Sin^2\left( \phi \right)  \end{matrix} \right] 
\end{equation} 
\begin{equation}
 Pr\left( Ay \right) = \left| \left| P_{By} | S \rangle \right| \right|^2 = \left( \left| Cos^2\left( \phi \right) \right| + \left| Sin^2\left( \phi \right) \right| \right) \left| \sqrt{s_0}Cos\left( \phi \right) + \sqrt{s_1} Sin\left( \phi \right) \right|^2
\end{equation}

\item To compute the probability of the sequence of answers $Ay \rightarrow By$, we need to compute the squared magnitude of their sequence of projections, that is $\left| \left| P_{By} P_{A_y} | S \rangle \right| \right|^2$:
\begin{equation}
Pr( By Ay ) = \left| \left| P_{By}~P_{A_y} | S \rangle \right| \right|^2 =  \left| s_0 \right| Cos^2\left( \phi \right)  \left(  Cos^2\left( \phi \right) + Sin^2\left( \phi \right) \right) 
\end{equation}

\item In the same way, we can compute the sequence of answers $Bn$ and $Ay$, representing the answer $yes$ to question $A$ and $no$ to question $B$.
\begin{equation}
Pr( Bn Ay ) =   \left| \left| P_{B_n}~P_{A_y} | S \rangle \right| \right|^2 = \left| Sin\left( \phi \right)  \left(- \sqrt{s_1} Cos\left( \phi \right)  + \sqrt{s_0} Sin\left( \phi \right) \right) \right|^2
\end{equation}

\item The final probability of $A$ being $yes$, $Pr(B \rightarrow A_y )$, is given by the sum: $Pr( B \rightarrow Ay) = Pr( By Ay ) + Pr( Bn Ay )$.
\begin{equation}
\begin{split}
Pr\left( B \rightarrow A_y \right) =Pr( B_y A_y ) + Pr( B_n A_y )=  
\left| Sin \left( \phi \right) \left( \sqrt{s_0}~Sin\left( \phi \right) - \sqrt{s_1}Cos \left( \phi \right) \right)  \right|^2 +\\
\left| Cos\left( \phi \right) \left( Cos\left( \phi \right)\sqrt{s_0} + Sin\left( \phi \right)\sqrt{s_1} \right)  \right| ^2~~
\end{split}
\label{eq:ay2}
 \end{equation} 

\item If, however, we want to compute the probability of $B_y$ in the second question, that is, the probability of answering $yes$ to $B$ after answering question $A$, then we would have to compute the sequence of projections:

\begin{equation}
 Pr( Ay By ) = \left| \left| P_{B_y}~P_{A_y} | S \rangle \right| \right|^2  = \left| s_0 \right| Cos^2\left( \phi \right) \left( Cos^2\left( \phi \right) + Sin\left( \phi \right)^2 \right) 
\end{equation}

\item The probability of answering $no$ to the first question $A$ and $yes$ to the second question $B$ is given by
\begin{equation}
 Pr( An By ) = \left| \left| P_{B_y}~P_{A_n} | S \rangle \right| \right|^2  = \left| s_1 \right| Sin^2\left( \phi \right) \left( Cos^2 \left( \phi \right) + Sin^2\left( \phi \right) \right) 
\end{equation}

\item The final probability of answering $yes$ to question $B$ followed by question $A$ is given by the sum of the probabilities $Pr( A \rightarrow By ) = Pr( AnBy) + Pr( Ay By )$
\begin{equation}
Pr \left( A \rightarrow By \right) = \left| s_0 \right| Cos^2\left( \phi \right) + \left| s_1 \right| Sin^2\left( \phi \right) 
 \label{eq:second_quest_final2}
\end{equation}

\item In the end, setting the parameters as suggested in~\cite{Busemeyer12book}, that is $\theta = \pi/4$, $s_0 = 0.7$ and $s_1 = 0.3$, this leads to a big gap in the non-comparative context and a small gap for the comparative context, simulating the {\it Assimilation Effect} presented in Table~\ref{tab:moores_data1}.

\[ Pr\left( Ay \right) = 0.7~~~~~Pr\left( By \right) = 0.96~~~~~Pr\left( A \rightarrow By \right) = Pr\left( B \rightarrow Ay \right) = 0.5 \]
\end{itemize}

It is important to note that, in this work, the main difference between a quantum-like model and a classical model resumes to the fact that in the first we use amplitudes (which are complex numbers) and in the later we use real numbers~\cite{busemeyer06}. The general idea is that, by using complex numbers, when we perform a measurement and make the squared magnitude of the projection, the usage of complex numbers will lead to the emergence of quantum interference effects. If, on the other hand, we use real numbers (classical model), then when we measure the length of the projection, we will never obtain any kind of interference terms. 

\subsection{Analysis of the Classical Projection Model}

In the quantum model, we concluded that the quantum interference term $\theta$ plays no role in a 2-dimensional model. In this section, we are interested to know if there is any relation between the classical model and the quantum model. In order to do this, we started to analyse the probability of answering $yes$ to $B$ in the second question, $Pr( A \rightarrow By )$. With the classical projection model one can achieve the same results reported in the quantum model. Since in the quantum model the quantum interference terms do not play any role in the computation of the probability of the second question (the one that has the rotated basis vectors), then it is straightforward that it can only accommodate the paradoxical finds through the rotation parameter $\phi$. In the same way, the classical model only depends on this rotation parameter in  order to simulate the several order effects reported in the work of Moore (2002)~\cite{Moore02}. 

We also decided to fix the rotation parameter $\phi$, in the classical model, and verify how the input state $s_0$ would affect the final probabilities. The analysis showed a constant function, which means that from the moment we specify a rotation to the model, it does not matter what the input state $s_0$ is, because it does not affect the final probability outcome. Just like in the quantum model, it is {\it only the rotation parameter} that will enable the accommodation of the several order effects reported in the work of Moore (2002)~\cite{Moore02}. We also made a similar analysis with respect the probability of answering $A$ in the second question. The results obtained reinforce the evidence that the quantum projection model has the same performance as the classical model.

\subsection{Explaining Serveral Order Effects using the Classical and Quantum Projection Models}

So far, we have seen that both classical and quantum models applied for the Clinton-Gore example give similar results. In this section, we will fit the different parameters of both models in order to simulate all order effects presented in the work of Moore (2002)~\cite{Moore02}. Table~\ref{tab:results} presents the results obtained. Since quantum interference terms did not play any role in the quantum model, we can see that the values used to fit the classical model are the same ones used to fit the quantum model. This reinforces the conclusion that, for a 2-dimensional decision scenario, the quantum projection model collapses to the classical model and does not provide any advantages towards the classical approach.

\begin{table}[ht]
\begin{center}
\resizebox{\columnwidth}{!} {
\begin{tabular}{c || c c c c c || c c c c }
\hline
 \multicolumn{1}{ c }{{~~~~~~~}} 	& \multicolumn{5}{ c }{{\bf ~~~~~~Quantum Projection Model~~~~~~}} 		& \multicolumn{4}{ c }{{\bf Classical Projection Model}}		\\   
	  \hline
 {\bf Experiments}  & {\bf $s_0$} & {\bf $\theta$} & {\bf $\phi$}	& {\bf Pr(1st ans) vs } & {\bf Pr(2nd ans) vs} & {\bf $s_0$} & {\bf $\phi$}	& {\bf Pr(1st ans) vs } & {\bf Pr(2nd ans) vs}~~~	\\
				&  	&	&	& {\bf Pr(1st ans exp)}	& {\bf  Pr(2nd ans exp)}	&	&	& {\bf Pr(1st ans exp)}	&  {\bf  Pr(2nd ans exp)}	\\
	
  \hline
{\bf Clinton / Gore}	& $\sqrt{0.50} e^{i \theta_0}$  & $\left[0, 2~\pi \right]$	& 0.7133	& 0.50 / 0.50	& 0.57 / 0.57	& $\sqrt{0.50}$	 & 0.7133		& 0.50 / 0.50 & 0.57 / 0.57 \\
{\bf Gore / Clinton}	& $\sqrt{0.68}e^{i \theta_0}$	& $\left[0, 2~\pi \right]$	& 2.6516	& 0.68 / 0.68	& 0.60 / 0.60 	& $\sqrt{0.68}$ &  2.6516	& 0.68 / 0.68 & 0.60 / 0.60  \\
\hline
{\bf Gingrich / Dole}		& $\sqrt{0.41} e^{i \theta_0}$	& $\left[0, 2~\pi \right]$ 	& 1.4858	& 0.41 / 0.41	& 0.33 / 0.33	& $\sqrt{0.41}$ & 1.4858		& 0.41 / 0.41	& 0.33 / 0.33 \\
{\bf Dole / Gingrich}		& $\sqrt{0.60}e^{i \theta_0}$	& $\left[0, 2~\pi \right]$	& $\pi$ 	& 0.60 / 0.60	& 0.60 / 0.64	& $\sqrt{0.60}$ & $\pi$		& 0.60 / 0.60	& 0.60 / 0.64 \\
\hline
{\bf While / Black}		&$\sqrt{0.41}e^{i \theta_0}$	& $\left[0, 2~\pi \right]$	& 0.0510	& 0.41 / 0.41	& 0.46 / 0.46	& $\sqrt{0.41}$	 & 0.0510		& 0.41 / 0.41 	& 0.46 / 0.46 \\
{\bf Black / White}		&$\sqrt{0.53}e^{i \theta_0}$	& $\left[0, 2~\pi \right]$	& $\pi$	& 0.53 / 0.53	& 0.53 / 0.56	& $\sqrt{0.53}$ & $\pi$		& 0.53 / 0.53	& 0.53 / 0.56 \\
\hline
{\bf Rose / Jackson}	&$\sqrt{0.64}e^{i \theta_0}$	& $\left[0, 2~\pi \right]$	& 3.0216	& 0.64 / 0.64	& 0.52 / 0.52	& $\sqrt{0.64}$ & 3.0216		& 0.64 / 0.64	& 0.52 / 0.52 \\
{\bf Jackson / Rose} 	&$\sqrt{0.45}e^{i \theta_0}$	& $\left[0, 2~\pi \right]$	& $\pi$	& 0.45 / 0.45	& 0.45 / 0.33	& $\sqrt{0.45}$	 & $\pi$		& 0.45 / 0.45	& 0.45 / 0.33 \\
\hline	
\end{tabular}
}
\end{center}
\caption{Prediction of the geometric approach using different $\phi$ rotation parameters to explain the different types of order effects reported in the work of~\cite{Moore02}. The columns {\emph Pr(1st ans) vs Pr(1st ans exp)} represent the answer to the first question obtained using the projection models and the value reported in~\cite{Moore02}, respectively. {\emph Pr(2nd ans) vs Pr(2nd ans exp)} represent the answer to the second question obtained using the projection models and the value reported in~\cite{Moore02}.  }
\label{tab:results}
\end{table}

From Table~\ref{tab:results}, one can also note that it was possible to fit all parameters of both models in order to accommodate the paradoxical findings reported for different order effects, namely assimilation, contrast, additive and subtractive effects. One can also note that, for the subtractive effects, in  the experiment regarding {\it Jackson / Rose}, the final probability computed achieved an error of $36\%$ when compared to the results reported in the work of Moore (2002)~\cite{Moore02}. This error occurred because there was no possible way to minimize Equation~\ref{eq:second_quest_final2} to such values. The most minimum value of the function was achieved by setting the rotation parameter $\phi$ to $\pi$. Despite this problem, in general, both classical and quantum projection approaches were proved to be similar for a 2-dimensional decision problem and were both able to accommodate all order effects.

In the next section, we will make a brief discussion on whether to use quantum models or classical models to accommodate order effects.

\subsection{Occam's Razor}

Given that in the quantum projection approach the quantum interference $\theta$ parameter ends up not playing any role in the calculations, the interference effects come only from the rotation of the basis (belief) vectors. In a true quantum setting, interference effects would emerge naturally due to the nature of complex numbers. For some event $\alpha \in A$ followed by a finite set of $N$ partition events $\beta_j \in B$, where $\alpha$ and $\beta_j$ are represented by complex numbers, the total probability of event $\alpha$ considering just two events, $N=2$, is given by Equation~\ref{eq:interf}. The resulting interference is different from the interference that is produced by rotating the belief vectors by $\phi$ radians.
\begin{equation}
\begin{split}
Pr( \alpha ) = \left| \sqrt{\beta_1} \sqrt{\alpha | \beta_1}~e^{i \theta_1} + \sqrt{\beta_2} \sqrt{\alpha | \beta_2}~e^{i \theta_2}\right|^2~=~~~~~~~~~~~~~~~~~~~~~~~~~~~~~~~~~~~~~~~~~~~~~~~~~~~\\
~=~ \left| \sqrt{\beta_1} \sqrt{\alpha | \beta_1}\right|^2 + \left| \sqrt{\beta_1} \sqrt{\alpha | \beta_1} \right|^2 + 2\cdot  \left| \sqrt{\beta_1} \sqrt{\alpha | \beta_1}\right| \left| \sqrt{\beta_2} \sqrt{\alpha | \beta_2} \right| Cos \left( \theta_1 - \theta_2 \right)
\end{split}
\label{eq:interf}
\end{equation}
For $N$ random variables, Equation~\ref{eq:interf} generalises to Equation~\ref{eq:interf_global}~\cite{Moreira16}, where one can notice an exponential growth of the quantum interference parameters $\theta$.
\begin{equation}
Pr( \alpha ) = \sum_{j=1}^N \left| \sqrt{\beta_j} \sqrt{\alpha | \beta_j} e^{i \theta_j} \right|^2 + 2 \sum_{j=1}^{N-1} \sum_{k=j+1}^N \left| \sqrt{\beta_j} \sqrt{\alpha | \beta_j} e^{i \theta_j} \right|  \left| \sqrt{\beta_k} \sqrt{\alpha | \beta_k} e^{i \theta_k} \right| \cdot Cos\left( \theta_j - \theta_k\right)
\label{eq:interf_global}
\end{equation}

For the case of order effects, interference terms end up not playing any role in order to determine the final probabilities in a decision scenario. However, this represents an exception. Many experiments have been conducted throughout the literature, which show that humans violate the laws of probability theory and logic under scenarios with high levels of uncertainty. Experiments such as The Prisoner’s Dilemma~\cite{Shafir92,Crosson99,Li02}, The Two Stage Gambling Game~\cite{Tversky92,Kuhberger01,Lambdin07} and categorization experiments~\cite{Busemeyer09markov} show that pure classical models cannot simulate human decisions, however using quantum probability, it is possible. The quantum interference terms that emerge from the application of complex numbers in the representation of mental states, gives rise to a free parameter that can be used to fit and explain these experiments. So far, the literature has proposed dynamical Quantum models~\cite{busemeyer06} based on Hamiltonian matrices and Schrödinger’s equation, Quantum-Like Bayesian Networks~\cite{Moreira16} that can explain both classical and quantum phenomena in a single model, and many more.

Summarising, both classical and quantum models are similar. Order effects can be explained by these two frameworks intuitively, since both models take advantage of the fact that matrix multiplications are non-commutative. One could argue that what makes the model quantum is the usage of these quantum amplitudes, which in turn generate quantum interference effects, which can be used to accommodate paradoxical situations, such as order effects. However, that does not hold in the quantum projection model, since quantum interferences end up playing no role in the computation of the probabilities. Moreover, even if the quantum interference terms did matter, then the complexity of the quantum projection model would increase, since we would require an additional $2^N$ free quantum interference parameters for binary questions that would need to be fit. To each input state $sx$ there is an additional phase parameter $e^{i\theta_x}$. For the case of $M$ possible answers, the complexity grows to $M^N$, where $N$ is the number of questions.

In the end, one can question, which approach should be used. We can either use a classical projection approach to simulate inconsistencies of order effects or we can use a quantum projection approach. According to Occam's razor, in the presence two competing hypothesis, the one that has the fewest assumptions (or the one that is simpler) should be chosen. This, of course is a very vague statement and we should take into account the nature of the problem itself and what we intend to prove with it. If we are merely interested in simulating the results of some experiments, abstracting ourselves from the interpretation and meaning of the nature of the experiment, then according to Occam's razor, the classical approach would be right choice. If, however, we are interested in developing a more general theory that requires additional interpretations, then the quantum approach would be the one to go with. So, in the end, the application of a Quantum or Classical approach depends on the nature of the problem and in what we intend to explain (or prove) with the application of the model. 

\section{Conclusions}\label{sec:conc}

Quantum probability theory has been gaining increasing attention in fields outside of physics~\cite{Moreira16survey}. Its general framework enables the representation of beliefs in a superposition state, which is a vector that comprises the occurrence of all possible beliefs. Moreover, this vector representation decouples itself from classical probability theory and it is not limited to the constraints of set theory. This means that empirical findings, such as order effects, can be easily explained by the non-commutativity of matrix operations under a quantum approach. In purely classical models, these order effects cannot be directly explained, because set theory is commutative.

In this work, we showed how one can take advantage of the geometric-based quantum theory in order to accommodate several order effects situations (additive, subtractive, assimilation and contrast effects) using the gallup reports collected in the work of Moore~\cite{Moore02}. However, we also showed that the exact same results can be obtained using a pure classical projection model. In the end, order effects can be explained by both frameworks intuitively, since both models are similar and take advantage of the fact that matrix multiplications are non-commutative. Depending on how one sets the rotation operator, one can simulate any effect reported in the work of Moore~\cite{Moore02}. Additionally, we also proposed an alternative interpretation for the rotation parameters used in these models, which is called the relativistic interpretation of parameters. This interpretation states that each person makes an inference by projecting a point in their personal $N$-dimensional psychological space, however, the person is not aware in which basis this point is projected in. So, the rotation of the parameters, instead of being interpreted as a measure of similarity (or an inner product) between questions, we state that this parameter emerges due to this lack of knowledge concerned with the basis state and also due to uncertainties towards the state of world and the context of the questions. With the relativistic interpretation of parameters, we can give to both classical and quantum approaches an interpretation for the rotation of the basis vectors and why this rotation is necessary under a cognitive point of view.

In the end, we argue that the application of these models should be based on Occam's Razor: in the presence two competing hypothesis, the one that has the fewest assumptions (or the one that is simpler) should be chosen. This depends much on the problem and what knowledge we want to extract from it. If we are mainly focused on a mathematical approach that can perform predictions for order effects, then the classical approach should be used. If, on the other hand, we want to make a model that leverages on theories and interpretations to explain the its predictions, then the quantum model is more indicated.

\bibliographystyle{spmpsci} 

\end{document}